\documentclass{emulateapj}
\usepackage{amssymb,bm}
\newcommand{\asca}{{\itshape ASCA}}
\newcommand{\chandra}{{\itshape Chandra}}
\newcommand{\xmm}{{\itshape XMM-Newton}}
\newcommand{\suzaku}{{\itshape Suzaku}}
\newcommand{\rate}{cts s$^{-1}$}
\newcommand{\lum}{ergs s$^{-1}$}
\newcommand{\flux}{ergs cm$^{-2}$ s$^{-1}$}
\newcommand{\surface}{ergs cm$^{-2}$ s$^{-1}$ sr$^{-1}$}
\newcommand{\edens}{ergs cm$^{-3}$}
\newcommand{\colden}{cm$^{-2}$}
\newcommand{\ue}{$u_{\rm e}$}
\newcommand{\um}{$u_{\rm m}$}
\newcommand{\src}{3C~326}
\newcommand{\HB}{[HB89]~1543+203}            
\newcommand{\WGA}{WARP~J1552.4+2007}         
\newcommand{\cluster}{WARP~J1552.2+2013}     

\slugcomment{}
\shorttitle{\suzaku~observation of \src}
\shortauthors{N. Isobe et al.}

\begin{document}
\title{\suzaku~Observation of the giant radio galaxy 3C 326}
\author{ 
Naoki      Isobe       \altaffilmark{1}, 
Makoto S.  Tashiro     \altaffilmark{2},
Poshak     Gandhi      \altaffilmark{3},
Asami      Hayato      \altaffilmark{3},
Hiroshi    Nagai       \altaffilmark{4},
Kazuhiro   Hada       \altaffilmark{5},
Hiromi     Seta        \altaffilmark{2},
Keiko      Matsuta     \altaffilmark{6}
}

\altaffiltext{1}{
        Department of Astronomy, Kyoto University, 
        Kitashirakawa-Oiwake-cho, Sakyo-ku, Kyoto 606-8502, Japan}
\email{n-isobe@kusastro.kyoto-u.ac.jp}
\altaffiltext{2}{
        Department of Physics, Saitama University,
        255 Shimo-Okubo, Sakura-ku, Saitama, 338-8570, Japan.}
\altaffiltext{3}{
        Cosmic Radiation Laboratory,
        the Institute of Physical and Chemical Research, 
        2-1 Hirosawa, Wako, Saitama, 351-0198, Japan}
\altaffiltext{4}{
  National Astronomical Observatory of Japan, 
  2-21-1 Osawa, Mitaka, Tokyo 181-8588, Japan}
\altaffiltext{5}{ 
        The Graduate University for Advanced Studies (SOKENDAI),
        2-21-1 Osawa, Mitaka, Tokyo 181-8588, Japan}
\altaffiltext{6}{ 
        The Graduate University for Advanced Studies (SOKENDAI),
        3-1-1 Yoshinodai, Sagamihara, Kanagawa, 229-8510, Japan}


\begin{abstract}
A \suzaku~observation of a giant radio galaxy, \src, 
which has a physical size of about $2$ Mpc, 
was conducted on 2008 January 19 -- 21.
In addition to several X-ray sources, 
diffuse emission was significantly detected associated with its west lobe,
but the east lobe was contaminated by an unidentified X-ray source \WGA.
After careful evaluation of the X-ray and Non X-ray background,
the $0.4$ -- $7$ keV X-ray spectrum of the west lobe is 
described by a power-law model modified with the Galactic absorption.  
The photon index and 1 keV flux density was derived as 
$\Gamma = 1.82_{-0.24}^{+0.26}\pm0.04$ and 
$S_{\rm X} = 19.4_{-3.2}^{+3.3}\pm 3.0$ nJy, respectively,
where the first and second errors represent the statistical and systematic ones.
The diffuse X-rays were attributed to be inverse Compton radiation 
by the synchrotron radio electrons 
scattering off the cosmic microwave background photons.
This radio galaxy is the largest among those with lobes 
detected through inverse Compton X-ray emission.
A comparison of the radio to X-ray fluxes
yields the energy densities of electron and magnetic field 
as 
$u_{\rm e} = (2.3 \pm 0.3 \pm 0.3) \times 10^{-13}$ \edens~and 
$u_{\rm m} = (1.2_{-0.1}^{+0.2}\pm 0.2 ) \times 10^{-14}$ \edens, respectively.
The galaxy is suggested to host a low luminosity nucleus 
with an absorption-corrected 2 -- 10 keV luminosity of $<2 \times 10^{42}$ \lum,
together with a relatively weak radio core. 
The energetics in the west lobe of \src~were compared with those of 
moderate radio galaxies with a size of $\sim 100$ kpc. 
The west lobe of \src~is confirmed to agree with the correlations 
for the moderate radio galaxies,
$u_{\rm e} \propto D^{-2.2\pm0.4}$ and $u_{\rm m} \propto D^{-2.4\pm0.4}$,
where $D$ is their total physical size. 
This implies that the lobes of \src~are still being energized by the jet, 
despite the current weakness of the nuclear activity.
\end{abstract}

\section{Introduction}
\label{sec:intro}
Lobes of many kinds of radio sources, including radio galaxies, 
are the energy storehouse on a large scale in the universe,  
of non-thermal electrons and magnetic fields,
both of which are conveyed from their active nuclei by the jets.
In the course of the progress of the jets,
the lobes expand with an estimated expansion velocity 
of about $0.01c$ -- $0.1 c$, where $c$ is the speed of light,
from a size of sub-kpc \citep[e.g.,][]{lobe_speed_CSO} 
to $\sim 100$ kpc \citep[e.g.,][]{lobe_speed}
on a time scale of about $1$ -- $10$ Myr.
Therefore, investigation of the energetics in lobes,
with respect to their size and/or age is thought to provide 
an important clue to probe into the history of the jet activity in the past. 

X-ray observations of inverse Compton (IC) emission 
from non-thermal electrons in lobes,
in which the cosmic microwave background (CMB) radiation 
plays role of a dominant seed photon source, are one of the ideal tools 
to examine the lobe energetics,  
because a comparison between the synchrotron radio and IC X-ray fluxes 
makes it possible to evaluate precisely the energy densities 
of electrons and magnetic fields, \ue~and \um~respectively \citep{CMB_IC}.
Motivated by the pioneering discoveries of lobe IC X-rays
with {\itshape ROSAT} and \asca~from Fornax A 
\citep{ForA_ROSAT,ForA_ASCA,ForA_ASCA_2}
and Centaurus B \citep{CenB}, several subsequent studies  
with \chandra~\citep[e.g.,][]{3C452,lobes_Croston,lobes_JK}
and \xmm~\citep[e.g.,][]{3C98,ForA,PicA_XMM} 
proved the usefulness of this technique. 
Moreover, such research has revealed 
an electron dominance of $u_{\rm e}/u_{\rm m} = 1$ -- $10$ 
typically in lobes of moderate radio galaxies
\citep[e.g.,][]{3C452,lobes_Croston}. 

The current IC results were limited to those from lobes of radio galaxies 
with a total size in a relatively narrow range of $50$ -- $500$ kpc. 
In order to investigate the progress of lobe energetics,
we urgently need to extend the size range of our sample. 
However, considering the angular resolution 
of the operating X-ray observatories 
(e.g., $\sim 0.5''$ for \chandra~
which corresponds to $\sim 1$ kpc at a redshift of $z \sim 0.1$),
it is thought to be not easy to observe smaller ones. 
Therefore, we focused on radio galaxies,
already evolved to more than $1$ Mpc.
These are called {\it giant radio galaxies} \citep[e.g.,][]{GRG0}.
Recently, 
\citet[e.g.,][]{3C457_Newton} reported the \xmm~result 
on the giant radio galaxy 3C 457, 
of which the lobes exhibit 
a magnetic field by a factor of $\sim 2$ weaker 
than that under the minimum energy assumption.

For such X-ray sources extended on a large angular scale 
(e.g., $ > 10'$ corresponding to $\sim 1$ Mpc at  $z \sim 0.1$),
\suzaku~\citep{Suzaku} has a great advantage over the other X-ray observatories,
thanks to its low and stable background \citep{xisnxbgen,HXD_NXB} and 
its large effective area up to more than $10$ keV.  
Actually, \citet{ForA_Suzaku} have confirmed IC X-ray emission from 
the west lobe of Fornax A,
in a wider energy range up to $\sim 20$ keV, with a higher accuracy. 

Located at the redshift of $z = 0.0895$ \citep{3C326_redshift},
\src~is a giant radio galaxy with an elliptical host \citep{3C326=Elliptical}. 
It is classified as a narrow line radio galaxy (NLRG). 
Its radio images \citep[e.g.,][]{GRG,GRG2} revealed a lobe-dominant 
Fanaroff-Riley (FR) II morphology \citep{FR}. 
The radio structure of \src~has a total angular size of $20'.1$, 
corresponding to $1.99$ Mpc at the source rest frame,
which makes the object one of the largest radio sources in the universe.
The object has a relatively high radio intensity of 
$7.4 \pm 0.1$ Jy at $609$ MHz \citep{GRG},
which ensures a high IC X-ray flux, 
in combination with its large physical size.  
These encouraged us to select this radio galaxy 
as a target for a \suzaku~observation. 

We adopted the cosmological constants of 
$H_{\rm 0} = 71$ km s$^{-1}$ Mpc$^{-1}$,
$\Omega_{\rm m} = 0.27$, and  $\Omega_{\lambda } = 0.73$.
In this cosmology, 
1 arcmin corresponds to $99.0$ kpc, at the redshift of \src. 

\begin{table}[b]
\caption{Best-fit spectral parameters of the source free region.}
\label{table:SRC_Free}
\centerline{\begin{tabular}{ll}
\hline \hline 
Parameter                                 & Value \\
\hline 
$N_{\rm H}$ ($10^{20}$ cm$^{-2}$)            & $3.84$ \tablenotemark{a} \\
$\Gamma $                                  & $1.41$ \tablenotemark{b} \\
$f_{\rm hard}$ (\surface) \tablenotemark{c} & $(6.0 \pm 0.5) \times 10^{-8}$          \\
$kT_{\rm 1}$ (keV)                         & $0.33 _{-0.07} ^{+0.21}$  \\
$kT_{\rm 2}$ (keV)                         & $0.11 _{-0.02} ^{+0.04}$  \\
$f_{\rm soft}$ (\surface) \tablenotemark{d} & $(2.5 \pm 0.2) \times 10^{-8}$\\
$\chi^2/{\rm d.o.f}$                       & $ 60.6 / 54$ \\
\hline 
\end{tabular}}
\tablenotetext{a}{Fixed at the Galactic value. }
\tablenotetext{b}{Taken from \cite{asca_cxb}. }
\tablenotetext{c}{The observed 2 -- 10 surface brightness 
                  of the PL component}
\tablenotetext{d}{The observed 0.5 -- 2 keV surface brightness 
                  of the sum of the 2 MEKAL components}
\end{table}

\begin{figure*}[t]
\plotone{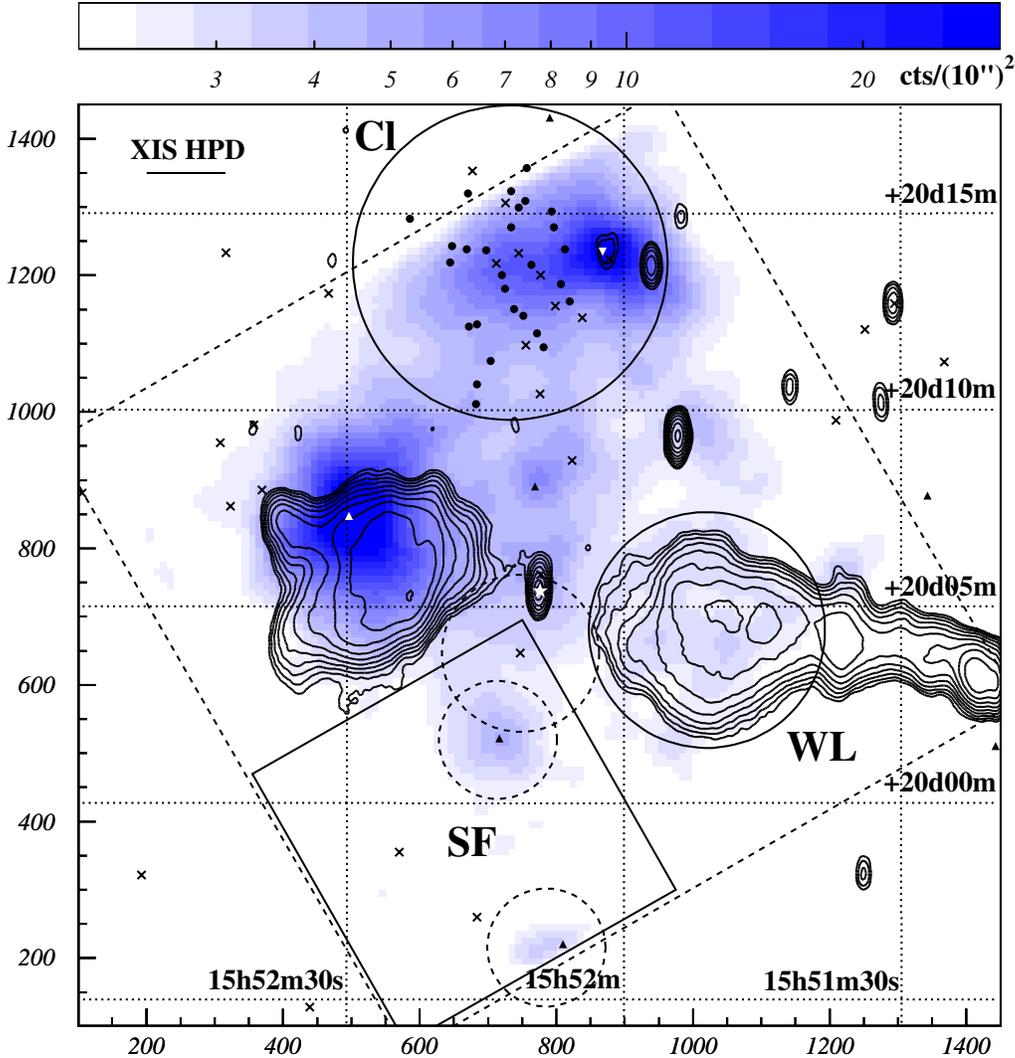}
\caption{
A blue-scale \suzaku~XIS image of 3C 326 in $0.5$ -- $5$ keV,
smoothed with a two-dimensional Gaussian function of a $20''$ radius.
Contours are overlaid from a 1.4 GHz radio image (Leahy et al. unpublished).
The scale bar on the top indicates 
the integrated X-ray counts within a $10'' \times 10''$ bin.  
The XIS FoV is shown with the dashed square.
The solid line at the top-left corner shows 
the point spread function of the XIS,
with a typical half power diameter (HPD) of $\sim 2$ arcmin \citep{XRTpaper}. 
The white filled star denotes the position of the optical host galaxy of \src. 
Member galaxies of the cluster \cluster~are referred by filled circles, 
and other galaxies by crosses. 
Black triangles corresponds to X-ray sources, taken from NED. 
The white normal and inverted triangles indicate 
the positions of \WGA~and \HB, respectively.
X-ray signals of the west lobe, source free region and 
the cluster \cluster~were 
accumulated from the regions denoted as {\bf WL}, {\bf SF} and {\bf Cl}, 
respectively. 
From the SF region, the dotted circles, corresponding to known X-ray sources, 
were removed.}
\label{fig:image}
\end{figure*}

\section{Observation and Data Reduction}
\label{sec:obs}
The \suzaku~observation of the giant radio galaxy \src~was
performed on 2008 January 19 -- 21.
The X-ray Imaging spectrometer \citep[XIS;][]{XISpaper}
and the Hard X-ray Detector \citep[HXD;][]{HXDdesign,HXDperform}
onboard \suzaku~were operated 
in the normal clocking mode without any window option, 
and in the normal mode, respectively.
The optical host galaxy of \src~was placed
at the XIS nominal position for the X-ray telescope 
\citep[XRT;][]{XRTpaper},
so that almost all the radio structure ($20'.1$) was 
within the XIS field of view (a $17'.8\times17'.8$ square).

We reduced and analysed the data 
with the standard software package, HEASOFT 6.5.1. 
All the data were reprocessed, referring to the CALDB as of 2008 July 9. 
In the following, we concentrate on the XIS data, 
since hard X-ray signals from the target were not detected 
significantly with the HXD. 
The following criteria were adopted for the data screening; 
the spacecraft is outside the south Atlantic anomaly (SAA),
the time after an exit from the SAA is larger than 436 s,
the geometric cut-off rigidity is higher than $6$ GV, 
the source elevation above the rim of bright and night Earth is 
higher than $20^\circ$ and $5^\circ$, respectively, 
and the XIS data are free from telemetry saturation. 
These procedures yielded about $50$ ks of good exposure.  
In the scientific analysis below,
we utilize only those events with a grade of 0, 2, 3, 4, or 6.

\section{Results}
\subsection{X-ray image}
\label{sec:image}
Figure \ref{fig:image} shows the 0.5 -- 5 keV \suzaku~XIS image 
of the \src~field, 
on which a 1.4 GHz radio contour image 
(Leahy et al. unpublished) is overlaid.
At the position of the host galaxy,  
(R.A., Dec) = ($15^{\rm h}52^{\rm m}09^{\rm s}.19$, $+20^{\circ}05' 23''.2$)
in the J2000.0 coordinate (the star in Figure \ref{fig:image}),
we detected no bright X-ray source,
while we found only a faint X-ray emission around it. 

The XIS field of view (FoV) is rather crowded 
with several contaminating X-ray sources.
The brightest point-like source detected at a J2000.0 coordinate of 
(R.A., Dec) = ($15^{\rm h}52^{\rm m}29^{\rm s}.02$, $+20^{\circ}07'16''.2$)
corresponds to a ROSAT source \WGA~\citep{WARP}.
No definite counterpart for the object has been 
reported at any other wavelength. 
We noticed another bright point-like source, 
a quasar \HB~with a redshift of $z = 0.25$ \citep{HB89_redshift},
at (R.A., Dec) = ($15^{\rm h} 52^{\rm m} 01^{\rm s}.61$, $+20^{\rm d} 13' 57''.5$). 
There is a cluster of galaxies,
\cluster~\citep{WARP} with a redshift of $z=0.136$ \citep{cluster_redshift},
on the north of \src,  
where a diffuse X-ray structure is clearly detected. 

In addition to these bright X-ray sources, 
a diffuse faint X-ray emission was detected from the west lobe of \src.
The spatial extent of the diffuse X-ray emission 
appears to be similar to that of the radio structure of the west lobe. 
On the contrary, severe contamination from \WGA~prevented us  
to investigate X-ray emission from the east lobe.

\subsection{X-ray spectrum of the source free region}
\label{sec:SRC_free}
Because the X-ray emission from the west lobe has 
a fairly low surface brightness, 
it is of crucial importance to evaluate precisely 
both X-ray and Non X-ray backgrounds (XRB and NXB, respectively). 
It is reported that the HEADAS tool {\tt xisnxbgen} 
reproduces the NXB spectrum with a systematic uncertainty of
better than $\sim 3$\% in the 1 -- 7 keV range, 
for a $50$ ks exposure \citep{xisnxbgen}. 

It is widely known that 
the XRB spectrum in the 0.2 -- 10 keV range is decomposed into 
a hard power-law (PL) component with a photon index of $\Gamma = 1.41$
and a two-temperature soft thermal plasma one \citep{asca_cxb,xmm_cxb}. 
The hard PL component is thought to 
be dominated by unresolved faint sources,
such as distant active galactic nuclei, 
and exhibits a small fluctuation in its flux 
\citep[$\lesssim 7$\%;][]{asca_cxb}.
On the other hand, 
the soft thermal component, 
which shows a significant field-to-field intensity variation, 
is thought to be associated with our Galaxy.

\begin{figure}[t]
\centerline{
\includegraphics[width=6.5cm,angle=-90]{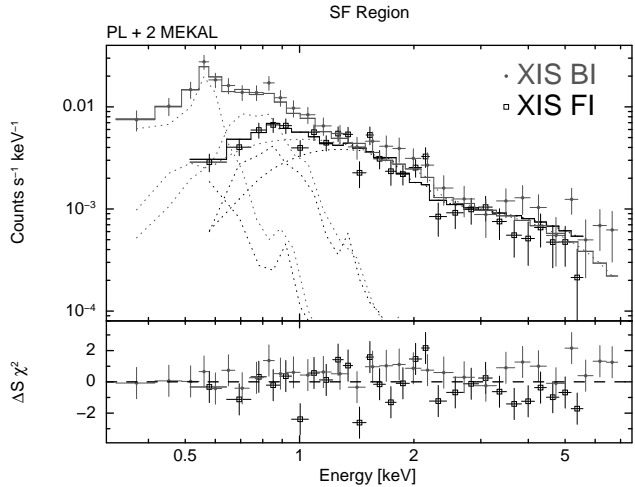}}
\caption{\suzaku~XIS spectrum of the SF region 
without removing instrumental response,
shown after the NXB events were subtracted.
The best-fit model,
consisting of a hard PL component and two soft MEKAL ones (dotted lines),   
is indicated by histograms in the top panel,
while the residuals to the model are displayed in the bottom panel.}
\label{fig:spec_SRC_Free}
\end{figure}

The XRB spectrum in this field
was extracted from the source free region 
within the FoV
(the square denoted as {\bf SF} in Figure  \ref{fig:image}),
without the known X-ray source regions (dashed circles). 
Figure \ref{fig:spec_SRC_Free} displays 
the XIS spectrum of the XRB, 
after the NXB, estimated by {\tt xisnxbgen}, was subtracted. 
Because 
the SF region is irradiated with a radioactive calibration source ($^{55}$Fe)
for the XIS FIs \citep{XISpaper}, 
we restricted the FI data below $5.5$ keV.
Consequently, 
the XRB signals were significantly detected in the 0.4 -- 5 keV range. 

Referring to the method in \citet{asca_cxb}, 
we approximated the shape of the observed XRB spectrum.
A composite model was adopted, 
consisting of a hard PL component and two soft MEKAL 
\citep[e.g.;][]{MEKAL} ones,
all of which were subjected to a photoelectric absorption
with the Galactic hydrogen column density 
\citep[$N_{\rm H} = 3.84 \times 10^{20}$ \colden;][]{NH}.
The photon index of the PL component was fixed at $\Gamma = 1.41$,
while the MEKAL temperatures were both left free. 
Because the result was found to be insensitive 
to the metalicity of the MEKAL components,
we adopted the solar abundance ratio.
We calculated response matrix functions (rmf) using {\tt xisrmfgen}. 
Auxiliary response files (arf) were generated 
by {\tt xissimarfgen} \citep{xissimarf}, 
assuming a diffuse source 
with a $20'$ radius and a flat surface brightness distribution. 
 
The model became acceptable 
($\chi^2/{\rm d.o.f} = 60.6/ 54$),
with the parameters summarised in Table \ref{table:SRC_Free}. 
The temperatures of the hot and cool MEKAL components 
are both reasonable \citep{xmm_cxb}, 
at $kT_{\rm 1} = 0.33_{-0.07}^{+0.21}$ keV
and $kT_{\rm 2} = 0.11_{-0.02}^{+0.04}$ keV, respectively.
The observed 0.5 -- 2 keV surface brightness 
of the sum of the thermal components 
was measured to be $f_{\rm soft} = (2.5 \pm 0.2) \times 10^{-9}$ \surface. 
We confirmed that the surface brightness of the hard PL component,
which was determined as 
$f_{\rm hard} = (6.0 \pm 0.5) \times 10^{-8}$ \surface~in the 2 -- 10 keV range, 
was consistent with the result of \citet{asca_cxb},
$f_{\rm hard} = (6.4 \pm 0.6) \times 10^{-8}$ \surface,
within the systematic field-to-field uncertainties
\citep[$\lesssim 7$\%;][]{asca_cxb}.

\subsection{X-ray spectrum of the west lobe}
\label{sec:west_lobe}
\begin{figure}[t]
\plotone{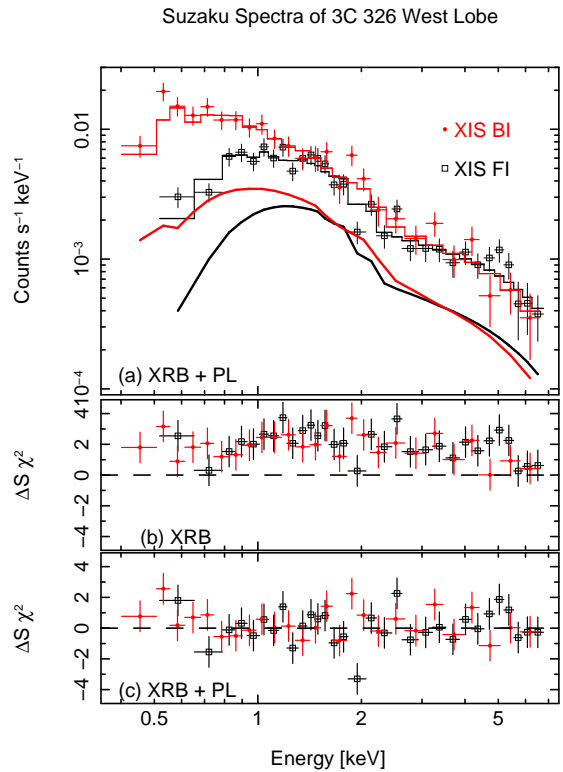}
\caption{\suzaku~XIS spectrum of the west lobe of 3C~326,
shown after only the NXB were subtracted. 
The histograms in Panel (a) show the best-fit XRB+PL model,
with the PL component indicated by the thick lines. 
Panels (b) and (c) show the residuals to the XRB model,
and those to the best-fit XRB+PL model, respectively. }
\label{fig:spec_west}
\end{figure}

The X-ray spectrum of the west lobe was integrated 
within the circle {\bf WL} in Figure \ref{fig:image},
with a radius of $3'$ (297 kpc in physical size).
In Figure \ref{fig:spec_west} (a),
we show it in the 0.4 -- 7 keV range, 
after subtracting only the NXB events 
in the manner similar to the X-ray spectrum of the SF region. 
The signal count rate over the NXB in the 0.4 -- 7 keV was measured 
as  $(1.22 \pm 0.04) \times 10^{-2}$ \rate~and 
$(1.88 \pm 0.07) \times 10^{-2}$ \rate~per CCD chip 
with the XIS FI and BI, respectively.

Adopting the best-fit XRB model spectrum determined from the SF region,
we estimated the XRB count rate to be 
$0.75 \times 10^{-2}$ cts s$^{-1}$ and $1.22 \times 10^{-2}$ cts s$^{-1}$,
within the WL region. 
Here, we precisely took into account 
the effect that the effective area to the XRB of the WL region
is $0.5$ -- $0.9$ times that of the SF region in the 0.5 -- 7 keV.
Thus, excess signals were significantly detected from the west lobe
above the XRB spectrum,
with a FI and BI count rate of $(0.47 \pm 0.04) \times 10^{-2}$ \rate~and 
$(0.66 \pm 0.07) \times 10^{-2}$ \rate, respectively.
The excess is clearly visualised in Figure \ref{fig:spec_west} (b),
which show the residual spectrum ($\chi^2/{\rm d.o.f} = 263.6 / 57 $)
of the west lobe over the XRB model.
The excess count rate is higher than 
the uncertainty of the XRB model ($\sim 8 \%$), derived in this observation. 

In order to evaluate the excess spectrum from the west lobe,
another PL component was introduced over the XRB model (XRB+PL). 
We generated the arf to the new PL component,
assuming the west lobe to be a spatially uniform X-ray source
with a $3'$ radius. 
Because the absorption column density to this component 
was unconstrained, we fixed it at the Galactic value. 
The additional PL component successfully reproduced the excess spectrum 
($\chi^2/{\rm d.o.f} = 61.9 / 55 $),
as shown in Figure \ref{fig:spec_west} (c). 
We show the best-fit PL component
with thick lines in Figure \ref{fig:spec_west} (a),
and its spectral parameters in Table \ref{table:west_lobe} (Case 1).
The photon index and flux density at 1 keV were determined to be 
$\Gamma = 1.82_{-0.24}^{+0.26}$ and $S_{\rm X} = 19.4_{-3.2}^{+3.3}$ nJy.
These correspond to a 0.5 -- 10 keV absorption-corrected 
X-ray flux of $F_{\rm X} = 1.7_{-0.2}^{+0.1} \times 10^{-13}$ \flux.

The uncertainty of the XRB model, obtained from the SF region, 
can affect the spectral parameters of the excess signals 
from the west lobe. 
Therefore, we re-analysed the X-ray spectrum of the WL region,
by changing the flux of the XRB model by $\pm 8$\%,
which corresponds to the uncertainty in our analysis 
(see Table \ref{table:SRC_Free}).  
As a result,
we evaluated the uncertainty in the 1 keV flux density of the PL component 
as $\Delta S_{\rm X} = 3.0$ nJy, 
while we found that its photon index stayed relatively unchanged 
with an error of $\Delta \Gamma =  0.04$. 

The west lobe could be contaminated 
by an extended X-ray emission from the cluster \cluster~on the north of 
\src~and/or other X-ray sources.
However, as shown in Figure \ref{fig:image},
there are no member galaxies of \cluster~nor known X-ray sources
within the west lobe. 
In order to investigate the possibility of 
the contamination from the cluster quantitatively,
we re-analysed the X-ray spectrum of the west lobe, 
by replacing the PL model with a MEKAL one,
of which temperature and metal abundance were both 
fixed at the best-fit value for the cluster emission
($kT = 4.1$ keV and $A = 0.58$, respectively; 
see \S\ref{sec:cluster} for the details). 
However the fit became significantly worse 
($\chi^2/{\rm d.o.f.} = 69.5 / 56$) 
in comparison with the PL model.
For the lobe spectrum, 
the PL model is supported rather than the cluster thermal emission,
at an F-test confidence level of $\sim 99$\%.
Therefore,
we safely ascribed the diffuse X-ray emission 
to that from the west lobe itself.

\begin{table}[t]
\caption{Spectral parameters of the west lobe of 3C 326}
\label{table:west_lobe}
\centerline{\begin{tabular}{lll}
\hline \hline 
                                   & Case 1                         & Case 2                   \\
\hline 
$N_{\rm H}$ ($10^{20}$ cm$^{-2}$)    & \multicolumn{2}{c}{$3.84$ \tablenotemark{a}}     \\
$\Gamma$  \tablenotemark{c}        & $1.82_{-0.24}^{+0.26} \pm 0.04$  & $ 1.8 $\tablenotemark{b}       \\
$S_{\rm X}$ (nJy) \tablenotemark{c} & $19.4_{-3.2}^{+3.3} \pm 3.0$    & $19.3 \pm 2.2 \pm 2.6 $             \\
$\chi^2/{\rm d.o.f}$               & $61.9 / 55$                    & $61.9 / 56$               \\
\hline 
\end{tabular}}
\tablenotetext{a}{Fixed at the Galactic value. }
\tablenotetext{b}{Fixed at the synchrotron radio index. }
\tablenotetext{c}{The first and second errors represent the statistical and systematic ones.}
\end{table}

\subsection{X-ray spectrum of the host galaxy of \src}
\label{sec:nucleus}
\begin{figure}[b]
\plotone{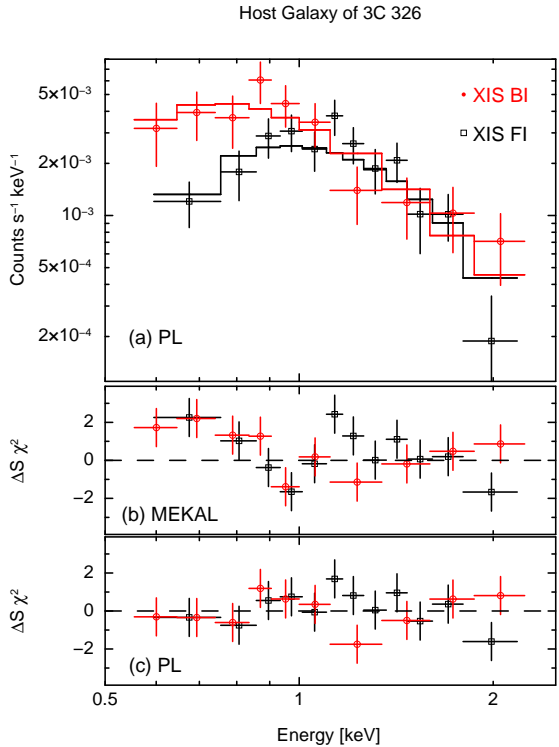}
\caption{\suzaku~XIS spectrum of the host galaxy of 3C 326.
The best-fit PL model is shown in Panel (a), with histograms. 
The residuals to the MEKAL and PL models are displayed 
in Panels (b) and (c), respectively. 
}
\label{fig:HostGalaxy}
\end{figure}

We derived X-ray signals associated with the host galaxy of 3C 326, 
including its nucleus, 
from a circle centered on it (the filled star in Figure \ref{fig:image}).
In order to avoid the contamination from the lobes 
and a bright unidentified X-ray source \WGA, 
a relatively small radius of $1.5'$ 
($149$ kpc at the source rest frame) was adopted.
As a sum of the NXB and XRB,
we simply subtracted the X-ray spectrum of the SF region, 
after rejecting that from the region corresponding to the calibration source.  
X-ray signals were significantly detected 
in the soft energy band with a 0.5 -- 2 keV count rate 
of $(2.4 \pm 0.2)\times 10^{-3}$ \rate~and 
$(3.3 \pm 0.4) \times 10^{-3}$ \rate~with the XIS FI and BI, respectively.  
On the other hand,
we found that the X-ray signals in the hard energy band
were marginal with a count rate of 
$(7.5\pm1.5) \times 10^{-4}$ \rate~and $(7.8\pm2.5)\times10^{-4}$
in the 3 -- 8 keV range. 
These indicate that the 3C 326 host galaxy exhibits 
a rather soft X-ray spectrum.

Figure \ref{fig:HostGalaxy} shows the \suzaku~XIS spectrum 
of the \src~host galaxy in the 0.5 -- 2 keV range. 
We tried to fit the spectrum with a MEKAL model,
modified by the Galactic absorption. 
The metal abundance was fixed at 0.4 solar, 
a typical value for nearby elliptical galaxies 
\citep{Abund_elliptical}.
The MEKAL model failed to reproduce the observed spectrum
($\chi^{2}/{\rm d.o.f} = 36.1/20$),
because of significant residuals below $\sim 1$ keV 
(the panel (b) in Figure \ref{fig:HostGalaxy}).
Although this may suggest an additional low-temperature component, 
we did not examine a two-temperature MEKAL model,
due to insufficient signal statistics. 
In order to estimate the flux and luminosity, 
we replaced the MEKAL model with 
a simple PL model modified by the Galactic absorption.
The PL model successfully approximated the data 
($\chi^{2}/{\rm d.o.f} = 15.8/20$),
with spectral parameters shown in Table \ref{table:host_galaxy}. 
The X-ray flux was measured to be 
$6.2 \times 10^{-14}$ \flux~in 0.5 -- 2 keV, 
which corresponds to an absorption-corrected luminosity 
of $1.6 \times 10^{42}$ \lum~at the redshift of 3C 326.

\begin{table}[t]
\caption{Spectral parameters of the host galaxy of 3C 326}
\label{table:host_galaxy}
\centerline{\begin{tabular}{lll}
\hline \hline 
Model                           & PL                      &  MEKAL \\ 
\hline 
$N_{\rm H}$ ($10^{20}$ cm$^{-2}$) & \multicolumn{2}{c}{$3.84$ \tablenotemark{a}  } \\
$\Gamma$ of $kT$ (keV)          & $3.38_{-0.34}^{+0.35}$    & $1.27_{-0.19}^{+0.25} $                   \\
Abundance (solar)               & ...                      & $0.4$                 \\
$F_{\rm X}$ (\flux)\tablenotemark{b} & $6.2\times10^{-14}$  & $4.0\times10^{-14}$                    \\
$L_{\rm X}$ (\lum) \tablenotemark{c} & $1.6\times10^{42}$   & $8.6\times10^{41}$                 \\
$\chi^2/{\rm d.o.f}$            & $15.8 / 20$             & $36.1 / 20$           \\
\hline 
\end{tabular}}
\tablenotetext{a}{Fixed at the Galactic value. }
\tablenotetext{b}{Observed flux in 0.5 -- 2 keV.}
\tablenotetext{c}{Absorption-corrected luminosity in 0.5 -- 2 keV at the source rest frame ($z=0.0895$).} 
\end{table}

\subsection{X-ray spectrum of contaminating sources}
\label{sec:sources}
As shown in Figure \ref{fig:image}, 
the XIS detected several contaminating X-ray sources. 
In the following, we briefly analyse their X-ray spectra,
although we regard detailed discussion on their nature 
to be beyond the scope of the present paper.

\subsubsection{The unidentified source \WGA}
\label{sec:1WGA}
\begin{figure}[t]
\centerline{
\includegraphics[width=6.5cm,angle=-90]{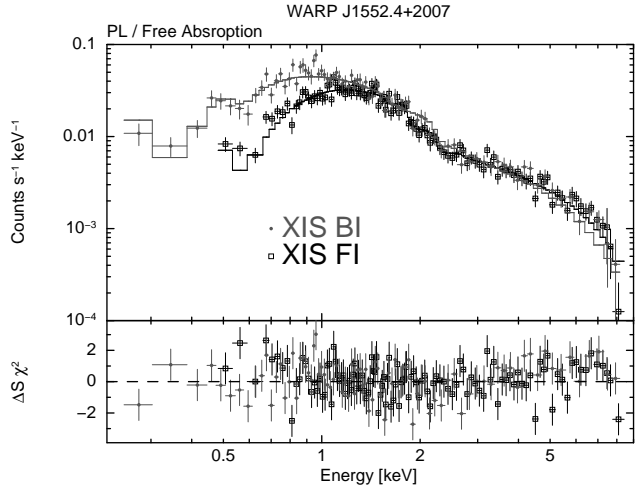}}
\caption{\suzaku~XIS spectrum of an unidentified X-ray source, \WGA,
shown with the best-fit PL model with histograms.}
\label{fig:1WGA_J1552.4+2007}
\end{figure}

The X-ray spectrum of an unidentified X-ray source \WGA~\citep{WARP}, 
the brightest source in this field, 
was extracted within the circle of a $2'.5$ arcmin radius, 
centered on the X-ray peak 
(the white triangle in Figure \ref{fig:image}). 
The XRB and NXB were subtracted, 
utilizing the events in the SF region,
with the calibration source region removed. 
Figure \ref{fig:1WGA_J1552.4+2007} shows 
the background-subtracted XIS spectra of the source,
in which the X-ray signals are significantly 
detected in the $0.3$ -- $8$ keV range.

Since no spectral features are seen, 
we examined a simple PL model modified by free absorption.
We obtained a reasonable fit ($\chi^2/{\rm d.o.f} = 233.7/197 $)
with parameters summarized in Table \ref{table:WGAandHB}. 
The fit suggests some residuals above $\gtrsim 6$ keV,
although we do not investigate this in detail. 
The absorption column density was consistent with the Galactic value.
The X-ray flux of \WGA~was measured 
as $F_{\rm X} = 8.6 \times 10^{-13}$ \flux~in 2 -- 10 keV.
The X-ray spectrum alone does not yield further insight 
into the nature of the source.

\begin{table}[b]
\caption{Summary of the PL fitting 
to the \suzaku~XIS spectra of \WGA~and \HB.}
\label{table:WGAandHB}
\begin{tabular}{lll}
\hline \hline 
Source                                 & \WGA                 & \HB                 \\ 
\hline 
$N_{\rm H}$($10^{20}$ \colden)           & $4.2_{-0.8}^{+0.9}$   & $6.3_{-2.5}^{+2.7}$    \\
$\Gamma$                               & $1.97_{-0.04}^{-0.05}$ & $2.13_{-0.10}^{+0.11} $    \\
$F_{\rm X}$ (\flux) \tablenotemark{a}   & $8.6 \times 10^{-13}$ & $4.4 \times 10^{-13} $    \\
$L_{\rm X}$ (\lum) \tablenotemark{b}    & ...                 &  $8.5 \times 10^{43}$    \\
$\chi^2/{\rm d.o.f}$                   & $233.7/197$         & $75.3 / 83$         \\ 
\hline 
\end{tabular}
\tablenotetext{a}{Absorption-inclusive 2 -- 10 keV flux}
\tablenotetext{b}{Absorption-corrected 2 -- 10 keV luminosity, at the source rest frame.}
\end{table}

\subsubsection{The quasar \HB}
\label{sec:HB89}
\begin{figure}[t]
\centerline{
\includegraphics[width=6.5cm,angle=-90]{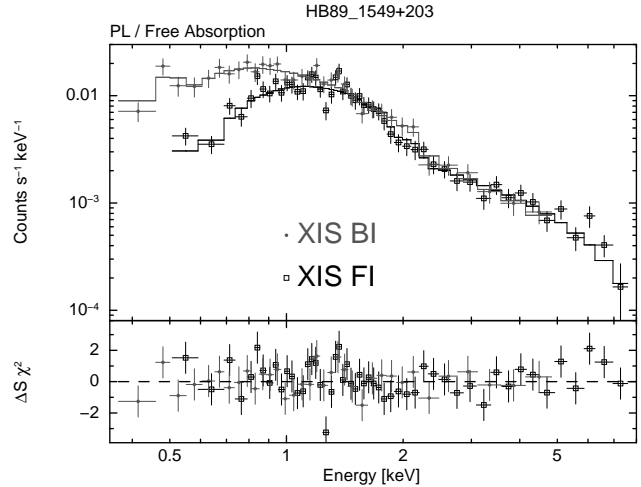}}
\caption{\suzaku~XIS spectrum of the quasar, \HB,
on which the best-fit PL model is overlaid.}
\label{fig:HB89_1543+203}
\end{figure}

In order to reduce the contamination from diffuse emission
associated with the cluster of galaxies \cluster,
a small circle of a $1'.5$ arcmin radius 
centered on \HB~(the white inverted triangle in Figure \ref{fig:image})
was adopted.
We subtracted the X-ray spectrum from the SF region,
in the manner similar to \WGA. 
Because the calibration source irradiates 
the top corner of the XIS BI, 
we limited the BI data to below 5 keV. 
As shown in Figure \ref{fig:HB89_1543+203},
the X-ray signals are significantly detected over $0.4$ -- $8$ keV.
 
We successfully reproduced the XIS spectrum of the quasar
($\chi^2/{\rm d.o.f} = 75.3 / 83 $)
by a simple PL model modified with an absorption,
whose column density is consistent with the Galactic value.  
The obtained parameters are listed in Table \ref{table:WGAandHB}.
The measured $2$ -- $10$ X-ray flux, 
$F_{\rm X} = 4.4 \times 10^{-13}$ \flux, 
corresponds to an X-ray luminosity of 
$L_{\rm X} = 8.5 \times 10^{43}$ \lum,
at the redshift of the quasar, $z = 0.25$. 
Using the {\itshape Einstein} observatory in 1981, 
the X-ray flux of \HB~was measured 
as $(1.37 \pm 0.11) \times 10^{-12}$ in the 0.3 -- 3.5 keV range, 
after the Galactic extinction was corrected \citep{EMSS}.  
Because the absorption-corrected \suzaku~flux 
in the same energy range was estimated to be $ 8.2 \times 10^{-13}$ \flux,
the source X-ray intensity has varied by a factor of $1.7$,
in 27 years between the {\itshape Einstein} and \suzaku~observations.

\begin{table}[b]
\caption{Spectral parameters to the \suzaku~XIS spectra of the cluster \cluster.}
\label{table:cluster}
\centerline{
\begin{tabular}{ll}
\hline \hline 
Parameter                                 & Value \\
\hline 
$N_{\rm H}$($10^{20}$ \colden)         & $3.84 $  \tablenotemark{a} \\ 
$kT$ (keV)                            & $4.1 \pm 0.5$     \\
$A$                                   & $0.58_{-0.28}^{+0.35}$   \\
$F_{\rm X}$ (\flux)\tablenotemark{b}   & $6.9 \times 10^{-13}$                  \\
$L_{\rm X}$ (\lum)\tablenotemark{c}    & $3.5 \times 10^{43}$                 \\
$\chi^2/{\rm d.o.f}$                  & $77.5/82$            \\
\hline 
\end{tabular}}
\tablenotetext{a}{Fixed at the Galactic value. }
\tablenotetext{b}{Absorption-inclusive 0.5 -- 10 keV flux. }
\tablenotetext{c}{Absorption-corrected 0.5 -- 10 keV luminosity, 
                  at the source rest frame ($z = 0.136$).}
\end{table}

\subsubsection{The cluster of galaxies \cluster}
\label{sec:cluster}
\begin{figure}[t]
\centerline{
\includegraphics[width=6.5cm,angle=-90]{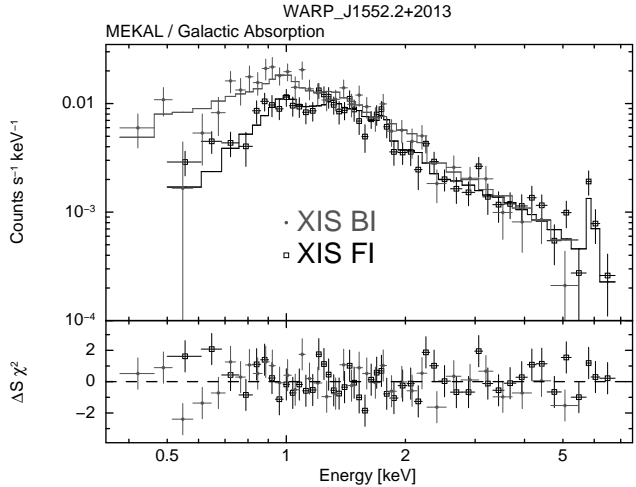}}
\caption{\suzaku~XIS spectrum of the cluster of galaxies, \cluster.
The histograms show the best-fit MEKAL model.}
\label{fig:cluster}
\end{figure}

As shown in Figure \ref{fig:image},
a number of member galaxies (filled circles) 
of the cluster \cluster~\citep{WARP} are 
within the XIS FoV, and associated diffuse emission is clearly seen. 
The X-ray spectrum of this cluster was taken from the circle denoted 
as {\bf Cl} in Figure \ref{fig:image}.
The circle of a $2'$ radius centered on \HB, 
and the region corresponding to the calibration source 
on the XIS BI were rejected from the Cl circle.
The SF region was adopted to subtract the NXB + XRB spectrum,
in the same way as for \WGA~and \HB.
The XIS spectrum of the cluster 
in the $0.4$ -- $7$ keV range is shown 
in Figure \ref{fig:cluster}.

We found an emission line feature around $\sim 5.8$ keV.
This energy, corresponding to $\sim 6.6$ keV 
at the rest frame of the cluster ($z = 0.136$),
is consistent with the energy of the ionized Fe emission lines. 
We, then, fitted the spectrum with a MEKAL model 
modified with the Galactic absorption ($\chi^2/{\rm d.o.f} = 77.5/82$),
and summarise the derived parameters in Table \ref{table:cluster}. 
The metal abundance ($A = 0.58_{-0.28}^{+0.35}$) and 
temperature ($kT = 4.1 \pm 0.5$ keV) 
are typical for cluster emission \citep{cluster_ASCA}. 
The X-ray luminosity of the cluster 
($L_{\rm X} = 3.5 \times 10^{43}$ \lum~in the 0.5 -- 10 keV range) 
may be slightly lower than that estimated 
from the luminosity-temperature relation of the typical clusters 
\citep[$L_{\rm X} \gtrsim 10^{44}$ \lum;][]{cluster_ASCA},
although we have to note that 
the XIS FoV did not cover the whole area of \cluster. 

\section{discussion}
\label{sec:discussion}
\subsection{Energetics in the west lobe}
\begin{figure}[t]
\plotone{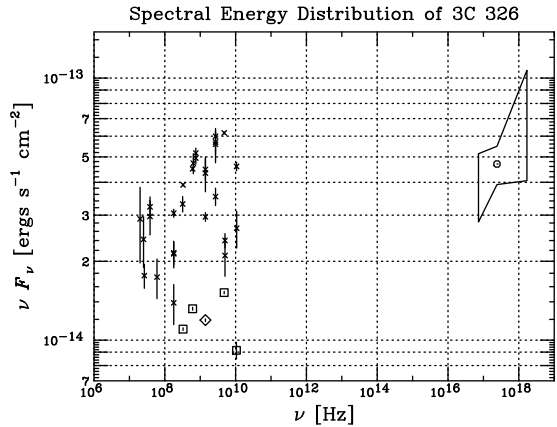}
\caption{Spectral energy distribution of 3C 326.
The best-fit PL model to the X-ray spectrum of the west lobe (Case 1) 
and its 1 keV flux density are plotted by the bow tie and circle, respectively.
Only the statistical errors are taken into account for the X-ray spectrum.
The crosses indicate the radio spectrum 
of the whole radio structure between 
20 MHz to 10.7 GHz
\citep{3C326_60MHz,
3C326_MHz,
3C_Catalogue,
3C326_10.7GHz,
RS_Catalogue,
GRG, 
radio_sources,
3C326_26.3MHz,
PKS_catalog}. 
The radio spectrum of the west lobe 
\citep[][the ``center'' region in their paper]{GRG2}
are shown with the boxes. 
The diamond presents the 1.4 GHz radio flux within the WL region,
which was evaluated from the radio image in Figure \ref{fig:image}.}
\label{fig:SED}
\end{figure}

We show the spectral energy distribution 
of \src~in Figure \ref{fig:SED} in $\nu F_{\rm \nu}$ units.
The radio spectrum of the west lobe is reported to exhibit 
a photon index of $\Gamma_{\rm R} = 1.7$ -- $1.9$ 
between 326 MHz and 4.8 GHz \citep[][]{GRG,GRG2}.
The X-ray photon index of the west lobe,
which we determined from the \suzaku~observation,
is consistent with this value.
This agreement strongly supports that 
the X-ray emission is produced by the synchrotron-emitting 
non-thermal electrons in the west lobe,
which approximately have a PL-like energy distribution. 
As is clear from Figure \ref{fig:SED},
the synchrotron spectrum of the west lobe shows 
a significant cut-off around 10 GHz, 
and is expected not to extend to the X-ray frequencies. 
Therefore, we safely attributed the diffuse X-ray emission 
from the west lobe to instead be of IC origin.
Like in the case of the southern inner lobe 
of Centaurus A \citep[e.g.,][]{CenA_shock},
the observed X-ray flux could be contaminated 
by synchrotron emission from high energy electrons 
accelerated in the shock region produced by the lobe expansion,
and thermal emission from the shocked and/or compressed plasma around the lobe.
However, 
a simple scaling of physical parameters from Centaurus A to 3C 326 by the size
assures that the X-ray flux of these components is far from detectable,
in this observation.

Among known radio galaxies with lobes detected through the IC X-ray emission,
\src~has the largest physical size ($\sim 2$ Mpc).
In lobes of radio galaxies on such a large scale ($\gg 100$~kpc)
with a low radio surface brightness, 
the seed photons of the IC process are though to be inevitably dominated 
by the CMB radiation, rather than other seed photon candidates.
These candidates include 
infra-red (IR) photons from the nucleus \citep[e.g.,][]{IC_nuclearIR}
with an energy density of $\lesssim 10^{-17}$ \edens,
estimated from the upper limit of the nuclear IR luminosity 
at a wavelength of $15~\mu$m of $\sim 1.4 \times 10^{42}$ \lum~\citep{RG_IR},
and synchrotron photons themselves with $\sim 10^{-18}$ \edens~from 
the SR radio flux determined below.
Actually, the CMB is evaluated to have a higher energy density 
of $u_{\rm CMB} = 5.8 \times 10^{-13}$ \edens,
at the rest frame of \src~($z = 0.0895$). 

\begin{table}[t]
\caption{Energetics in the west lobe of 3C 326}
\label{table:energetics}
\centerline{
\begin{tabular}{lcccc}
\hline\hline 
                       &                          & \multicolumn{3}{c}{Systematic errors from} \\ 
Parameters             & Value\tablenotemark{a}   & $S_{\rm R}$     & $\Gamma_{\rm R}$   &$V$\\
\hline       
\ue~($10^{-13}$\edens)  & $2.3\pm0.3 \pm0.3$       & $\pm0$          & $\pm0.3$         & $_{-0.4}^{+0.6}$         \\  
\um~($10^{-14}$\edens)  & $1.2_{-0.1}^{+0.2} \pm0.2$ & $_{-0.1}^{+0.2}$  & $_{-0.4}^{+0.5}$   & $\pm 0 $         \\  
$B$ ($\mu$G)           & $0.55_{-0.03-0.04}^{+0.04+0.05}$
                                                  & $_{-0.04}^{+0.03}$ & $_{-0.09}^{+0.10}$ & $\pm 0 $         \\  
$u_{\rm e} / u_{\rm m}$  & $18.8_{-4.3-4.9}^{+4.9+5.7} $   
                                                  & $_{-2.1}^{+2.6}$  & $_{-7.0}^{+12.2}$  & $_{-3.5}^{+4.7}$          \\  
\hline       
\end{tabular}}
\tablenotetext{a}{The first and second errors represent those 
                  propagated from the statistical and systematic ones in $S_{\rm X}$, respectively. }
\end{table}

We diagnose the energetics in the west lobe of \src,
by a comparison between the synchrotron radio and IC X-ray spectra. 
By examining the linear radio brightness profile 
from the 1.4 GHz radio map shown in Figure \ref{fig:image},
we approximated the shape of the lobe at a sphere with a radius of 
$140'' \pm 10''$ after deconvolving the beam size 
($39'' \times 14''$ in a full width at half maximum). 
This corresponds to a physical size of $231\pm23$ kpc, 
which gives a volume of $V=(1.5\pm0.3) \times 10^{72}$ cm$^3$. 
We estimated the radio intensity within 
the X-ray integration region of the west lobe (WL in Figure \ref{fig:image})
to be $S_{\rm R} = 0.85 \pm 0.09$ Jy at 1.4 GHz, from the radio image.
This flux is shown with the diamond in Figure \ref{fig:SED}. 
Based on the radio data below the GHz range \citep[e.g.,][]{GRG,GRG2},
we adopted the radio photon index of $\Gamma_{\rm R} = 1.8$ 
with an error of about $ \pm 0.05$,
which corresponds to the slope of the electron number density spectrum 
to be $p = 2\Gamma_{\rm R} - 1 = 2.6 \pm 0.1 $. 
In consequence, we re-evaluated the X-ray flux density 
with the photon index fixed at $\Gamma_{\rm R} = 1.8$,
as $S_{\rm X} = 19.3 \pm 2.2 \pm 2.6$ nJy at 1 keV
(Case 2 in Table \ref{table:west_lobe}). 
We supposed that the magnetic fields are randomly oriented over the lobe. 
The filling factors of the electrons and magnetic field were
assumed to be unity. 

We simply referred to \citet{CMB_IC},
and evaluated the energetics in the west lobe 
as summarised in Table \ref{table:energetics}.
The energy densities of electrons and magnetic field are calculated 
as $u_{\rm e} = (2.3 \pm 0.3 \pm 0.3 ) \times 10^{-13}$ \edens~and 
$u_{\rm m} = (1.2_{-0.1}^{+0.2} \pm 0.2) \times 10^{-14}$ \edens, 
respectively.
Here, \ue~was evaluated for the electrons 
with a Lorenz factor of $\gamma_{\rm e} = 10^3$ -- $10^{5}$,
because they are directly visible
through the synchrotron radio and/or IC X-ray emission.
The first and second errors are 
due to the statistical and systematic ones of the X-ray flux density, 
$S_{\rm X}$.  
We found an electron dominance of 
$u_{\rm e}/u_{\rm m} = 18.8_{-4.3-4.9}^{+4.9+5.7} $.
As a result, the magnetic field,
$B = 0.55 _{-0.03 -0.04} ^{+0.04 +0.05}$ $\mu$G,
was slightly weaker than that evaluated 
under the minimum energy condition \citep[e.g.;][]{Bme} 
of $B_{\rm me} \gtrsim 1.1$  $\mu$G, neglecting the proton contribution.
Even though all the possible systematics,
compiled in Table \ref{table:energetics},
are considered,
the electron dominance in the west lobe of \src~is justified.

\subsection{Low luminosity nucleus in 3C 326}
\label{sec:nucleus}
Recent studies with \chandra~and \xmm~
\citep[e.g.,][]{RG_nucleus1,RG_nucleus2,RG_nucleus3,RG_IR} revealed
that nuclei of FR II NLRGs exhibit both soft and hard spectral components,
the latter of which are heavily absorbed with a column density of 
$N_{\rm H} \gtrsim 10^{23}$ \colden. 
Adopting a PL spectrum with a photon index of $\Gamma = 1.7$
subjected to a column density of $N_{\rm H} = 10^{23}$ 
\colden~(typical values for  NLRG nuclei),
the marginal signals above 2 keV placed an upper limit 
on the absorption-corrected 2 -- 10 keV X-ray luminosity of the \src~nucleus 
as $2 \times 10^{42}$ \lum.
Because this upper limit is lower than the typical luminosity of NLRG
nuclei by an order of magnitude,
\citep[$\gtrsim 10^{43}$ \lum,][]{RG_nucleus1,RG_nucleus2,RG_nucleus3},
the activity of \src~is suggested to be relatively weak, among NLRGs.

In terms of the de-absorbed X-ray luminosity,
the soft X-ray component from NLRG nuclei 
is found to be dominated by the hard obscured one,
typically by an order of magnitude 
\citep{RG_nucleus1,RG_nucleus2,RG_nucleus3}.
In the case of \src,
the upper limit on the hard absorbed component luminosity 
corresponds to the contribution of the nuclear soft X-ray emission 
to be less than $\sim 10^{41}$ \lum. 
The observed soft X-ray luminosity from the \src~host galaxy,
$1.6 \times 10^{42}$ \lum, highly exceeds this value.
In addition, the 0.5 -- 2 keV X-ray spectrum of \src~with
a photon index of $\Gamma = 3.38_{-0.34}^{+0.35}$
is softer than those of the soft component in NLRG nuclei, $\Gamma \sim 2$ 
\citep[at most $\Gamma < 3$,][]{RG_nucleus1,RG_nucleus2,RG_nucleus3}.
These indicate that a large part of the soft emission 
observed from the \src~host galaxy is not of nuclear origin. 
Because the soft component from NLRG nuclei
is considered to originate in jets \citep{RG_ROSAT},
it is implied that the jet from \src~appears to exhibit only a weak activity,
although other origins \citep[e.g.,][]{3C33_softcomp}
for the soft component are not yet ruled out.

The relatively large integration circle 
with a radius of 149 kpc at the rest frame of \src,
due to the \suzaku~angular resolution, inevitably allows 
a significant contribution from the thermal emission 
associated with the host galaxy.
Actually, the observed X-ray luminosity,  $1.6 \times 10^{42}$ \lum,
is compatible with those of the thermal plasma
in nearby elliptical galaxies \citep{Abund_elliptical}. 
Therefore, we attributed the soft X-ray emission from \src~
to be dominated by a thermal plasma emission from the host galaxy,
although its companion galaxy,
which locates $\sim 25"$ north of \src~\citep[$z=0.0885$;][]{OpticalID_3C326},
may contribute to some extent. 

In order to reinforce the low luminosity nucleus in \src,
suggested by the \suzaku~observation,
we show other observational supports in the following.  
The host galaxy of \src~is reported to exhibit
no significant emission line of [O{\sc iii}] 
(at a wavelength of $\lambda = 5007$ \AA) 
with an upper limit of $\sim2\times10^{40}$ \lum \citep{3C326_emission_line}.
This emission line is regarded 
as one of the important signatures of activity of galactic nuclei. 
Using the correlation between the X-ray and  [O{\sc iii}]  luminosities
for nearby Seyfert galaxies \citep{LX-OIII},  
the upper limit on the nuclear X-ray luminosity of \src~is estimated 
to be $L_{\rm X}\sim 1 \times 10^{42}$ \lum~in the $2$ -- $10$ keV range.
Recently, \citet{RG_IR} reported 
another evidence for inactivity of the \src~nucleus from IR observations.
They measured an upper limit on its IR luminosity to be $\sim 1.4 \times 10^{42}$ \lum, 
at a wavelength of $15~\mu$m. 
The correlation between the X-ray and IR luminosities,
recently found for local Seyferts \citep{IR-X-Seyferts}, 
translates this into an upper limit on the $2$ -- $10$ keV X-ray luminosity  
of $L_{\rm X} \sim ~ 1 \times 10^{42}$ \lum.
All of these are consistent with our result. 

Figure \ref{fig:image} displays a radio source 
at the center of the \src~host galaxy. 
However, the radio core of \src~is reported to exhibit 
a rather low 5~GHz luminosity ($L_{\rm 5 GHz} \sim 1 \times 10^{40}$ \lum)
among NLRGs 
\citep[$L_{\rm 5 GHz} = 10^{39}$ -- $10^{42}$ \lum;][]{RG_ROSAT,RG_IR}. 
A correlation was reported for NLRGs
between the 178 MHz luminosity of the whole radio structure and 
the X-ray luminosity of the nuclear hard component \citep{RG_nucleus3,RG_IR}. 
Adopting this correlation,
the 178 MHz luminosity of \src~\citep[$\sim 8 \times 10^{41}$ \lum,][]{RG_IR} 
gives an estimate to 
the $2$ -- $10$ keV absorption-corrected luminosity as $\sim 10^{43}$ \lum,
which is a factor of 10 larger than the derived upper limit. 
Thus, the nucleus of \src, including its jets, 
seems less active for its radio lobe luminosity.
Because the extended lobes 
trace the past activity of the jet emanating from the nucleus, 
the \src~nucleus is considered to have been more active in the past, 
and to become relatively weak, recently.
A similar decline in the nuclear activity was reported from 
the radio galaxy Fornax A \citep{ForA_nucleus},
with a current $0.3$ -- $8$ keV luminosity of 
$L_{\rm X} = 5 \times 10^{39}$ \lum~\citep{ForA_chandra}.

\subsection{Evolution of lobe energetics}
\begin{figure}[t]
\plotone{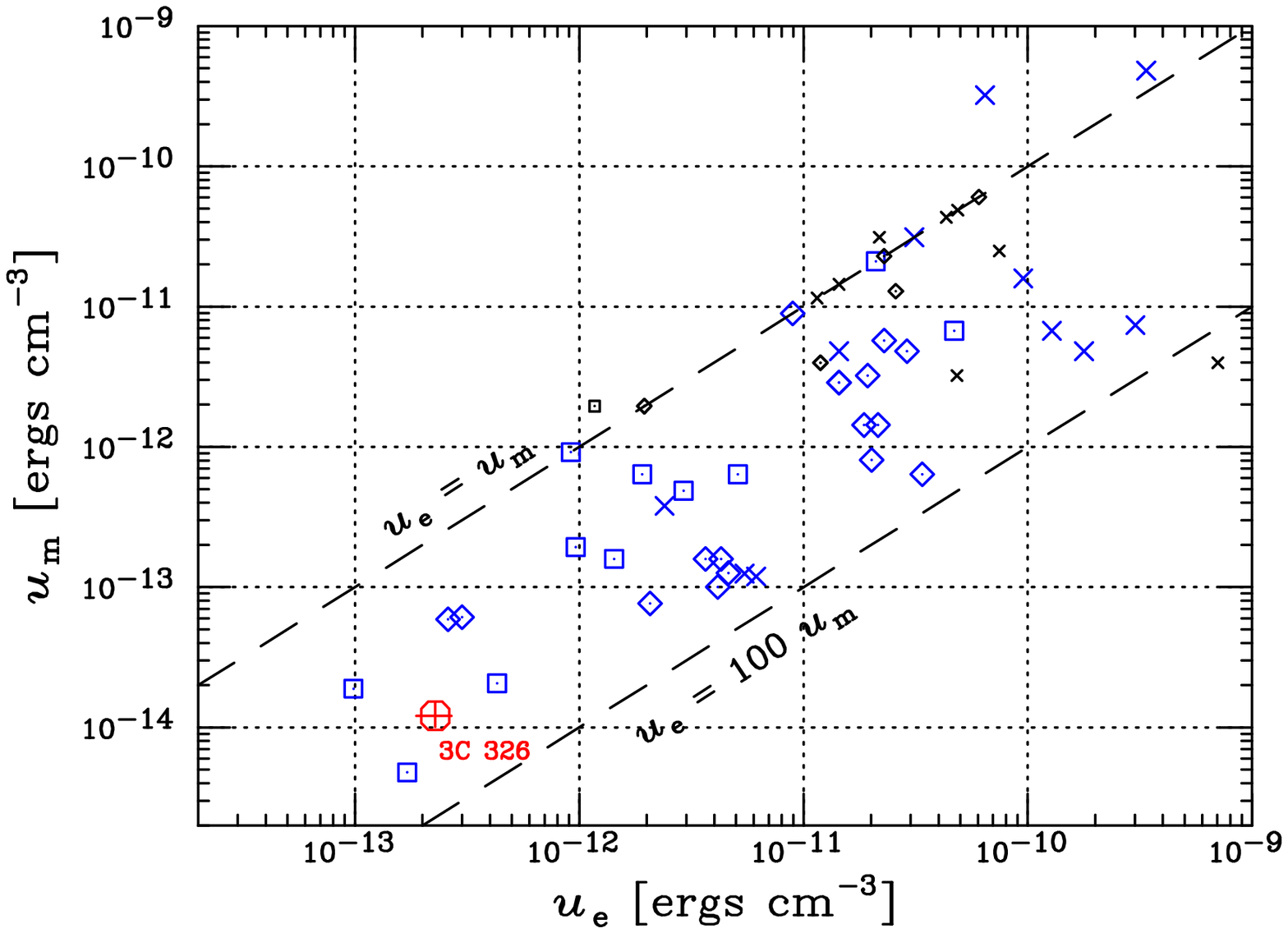}
\caption{The relation between \ue~and \um, 
determined through the IC technique 
\citep{lobes_Croston,Isobe_D,3C452,3C98,ForA,CenB,ForA_Suzaku}.
For the result on the west lobe of \src~shown with the red circle,
the propagation from the statistical and systematic errors 
in $S_{\rm X}$ are considered. 
Crosses, diamonds, and squares corresponds to lobes of radio galaxies,
with $D<200$ kpc, $D = 200$ -- $400$ kpc, and $D> 400$ kpc, respectively. 
Lobes, from which only the upper limit of IC X-ray flux was obtained, 
are indicated by the small black points, 
while the others are shown as the large blue ones. 
The two dashed lines represents $u_{\rm e} / u_{\rm m} = 1 $ and $100$. 
}
\label{fig:ue-um}
\end{figure}

In order to investigate evolution of the lobe energetics, 
we compared \src~with other radio galaxies accompanied by smaller lobes. 
Figure \ref{fig:ue-um} summarizes \ue~and \um~in lobes of radio galaxies,
determined through IC X-ray technique
\citep{lobes_Croston,Isobe_D,3C452,3C98,ForA,CenB,ForA_Suzaku}.
We confirmed that 
almost all the radio lobe are distributed between the lines of  
$u_{\rm e} / u_{\rm m} = 1$ and $100$ 
\citep[e.g.,][]{3C98,lobes_Croston}, over a wide range of \ue~and \um.  
The west lobe of \src~fully follows this trend. 
Moreover, Figure  \ref{fig:ue-um} indicates that 
radio galaxies on a larger scale tend to exhibit 
a smaller value of \ue~and \um~in their lobes,
while the degree of electron dominance $u_{\rm e} / u_{\rm m}$ seems 
to be independent of the size.

\begin{figure*}[t]
\plottwo
{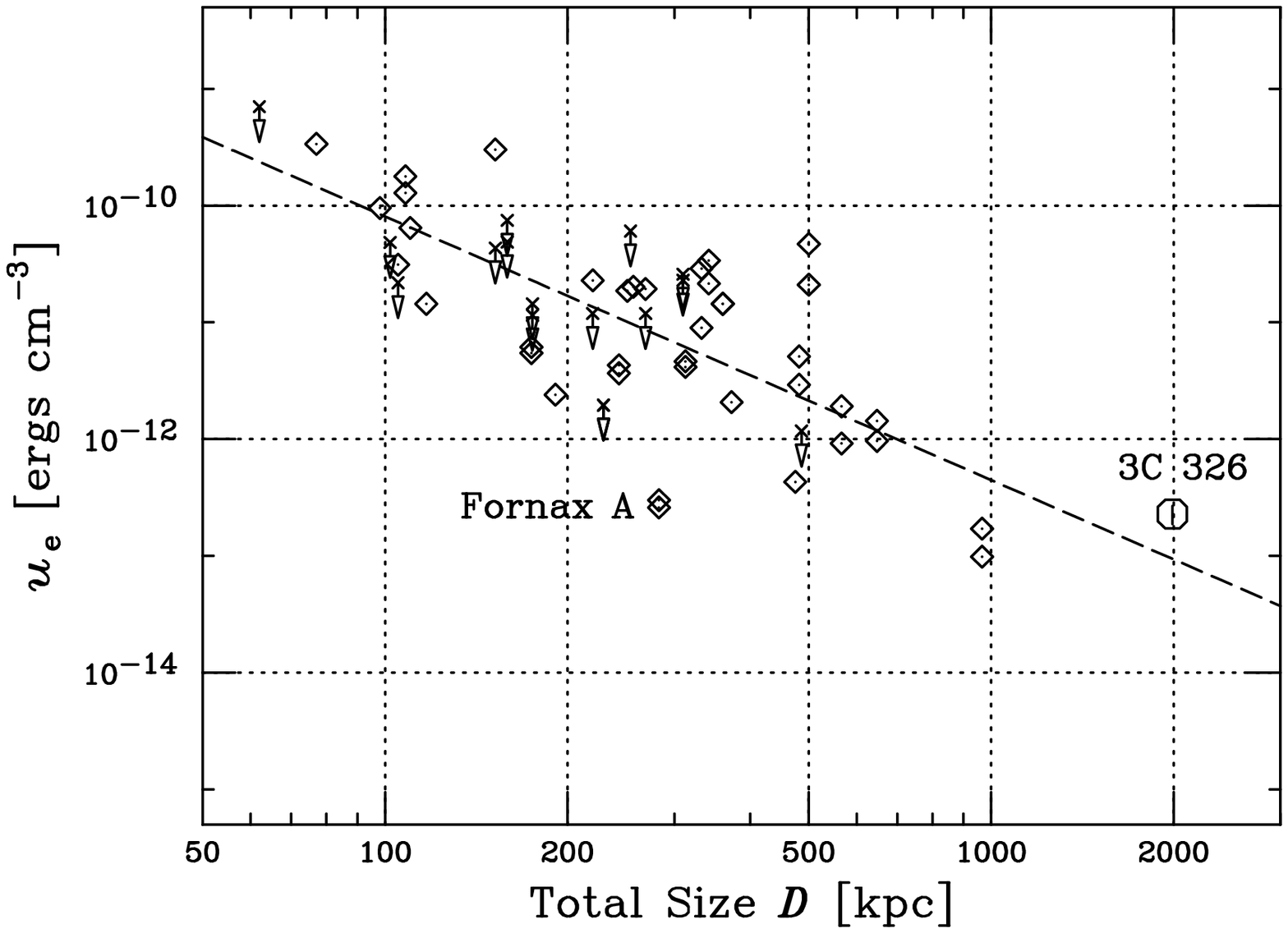}{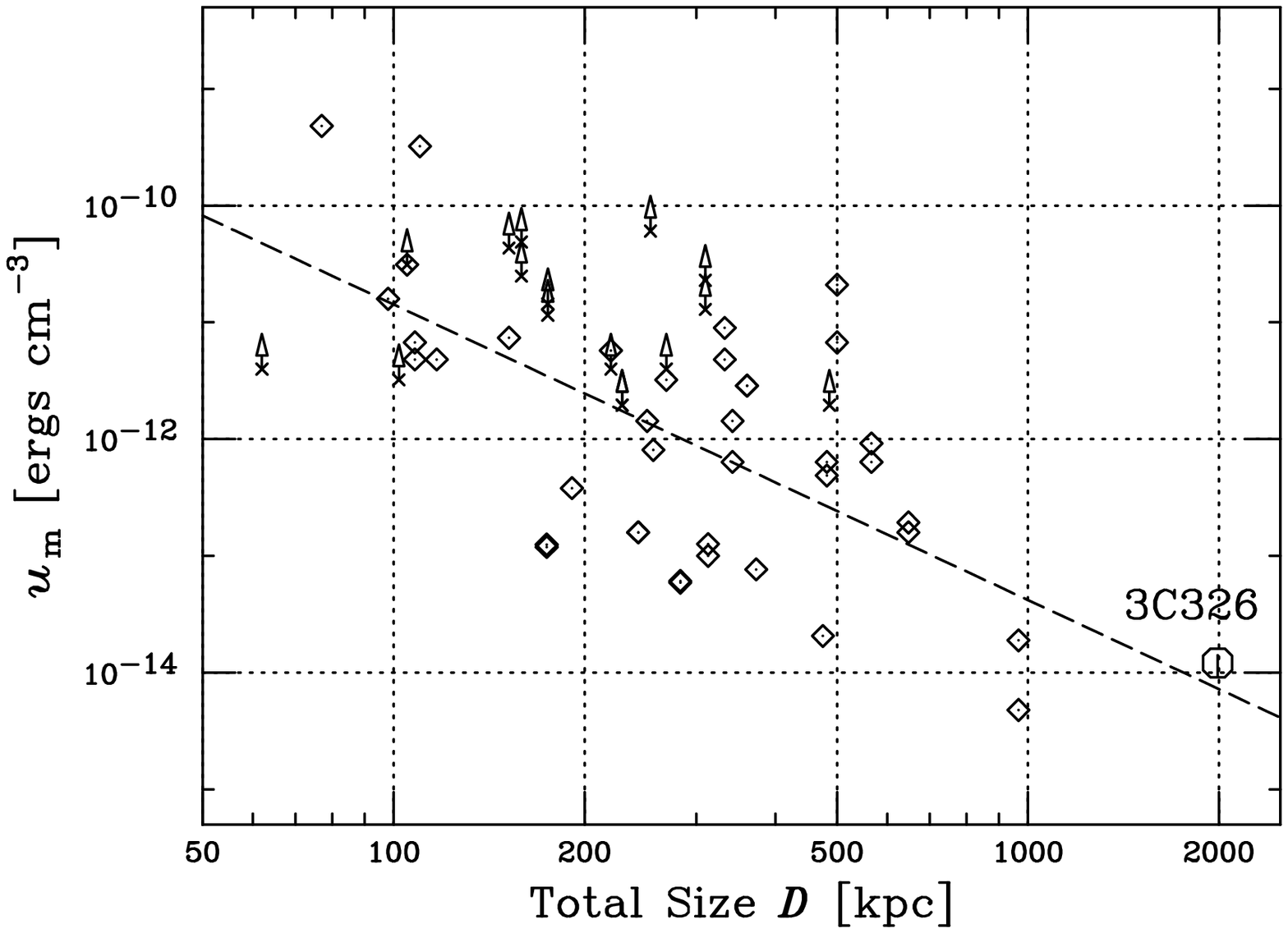}
\caption{The energy densities of electrons (left) and magnetic fields (right),
\ue~and \um~respectively, in lobes
\citep{lobes_Croston,Isobe_D,3C452,3C98,ForA,CenB,ForA_Suzaku},
plotted against the total size, $D$, of radio galaxies. 
The circle corresponds to the west lobe of \src, 
for which the statistical and systematic errors in $S_{\rm X}$ 
are taken into account.
Diamonds shows lobes from which the IC X-rays were securely detected,
while crosses indicate those with only the upper limit on the IC flux. 
The dashed lines represent the best-fit relations  
of $u_{\rm e} \propto D^{-2.2}$ and $u_{\rm m} \propto D^{-2.4}$, 
for lobes displayed with diamonds. }
\label{fig:size2ue}
\end{figure*}

The dependence of \ue~and \um~on the total dimension $D$ 
of the radio galaxies 
is more clearly visualised in Figure \ref{fig:size2ue}.
Recently, \citet{size2age} reported a trend 
that the total size of radio sources
is almost proportional to the source age ($\tau = 100$ yr --  $10$ Myr),
estimated by the synchrotron aging technique,
over a wide range of $D = 10$ pc -- $1$ Mpc.
This proportionality is thought to be justified 
by the fact that the observed expansion speed of lobes 
are distributed in a relatively narrow range of $0.01c$ -- $0.1 c$
\citep[e.g.,][]{lobe_speed,lobe_speed_CSO}.
Thus, the figure can be regarded as a plot between \ue, \um~and $\tau$. 
For lobes with $D < 1$ Mpc,
we found relations of 
$u_{\rm e} \propto D^{-2.2 \pm 0.4}$ and $u_{\rm m} \propto D^{-2.4 \pm 0.4}$ 
(the dashed lines in Figure \ref{fig:size2ue}),
although the data points have a slightly large scatter 
(about less than an order of magnitude) 
around the correlation line. 
In order to derive the relation, 
we ignored the lobes in which only the upper limit of \ue~
(and hence the lower limit of \um) was derived. 
The data point of the giant radio galaxy \src~($D = 1.99$ Mpc) seems 
to agree with this tendency for the smaller source with $D<1$ Mpc,
within a factor of 2.  

The observed correlation could be an artifact,
due to the inclination angle, $\theta$, 
of the radio source axis to the sky plane.
As $\theta$ increase, 
the apparent dimension and depth of non-spherical sources 
becomes shorter and longer, respectively.
In addition, the longer depth of the emitting region 
can enhance the X-ray and radio surface brightness.
However, we think that these effects have no impact on 
the spatially averaged magnetic field strength $B$ and its energy density \um,
because they are determined from the radio and X-ray fluxes 
spatially integrated over the region of interest, 
$S_{\rm R}$ and $S_{\rm X}$ respectively,
where the $\theta$ dependence is canceled out.
The spatially averaged electron energy density \ue~is proportional 
to the observed value of $S_{\rm X} V^{-1}$.  
Because the apparent volume of the source 
linearly depends on its apparent dimension,
we expect that $\theta$ is possible to 
cause a fake dependence of $u_{\rm e} \propto D^{-1}$.
However, this imitation can not explain the observed relation.
Therefore, we regard the observed correlation 
between \ue, \um and $D$ to be real. 

What does the relation tell us?
For simplicity, we assume that the energy input rate by jets to lobes 
is common among radio galaxies, and constant in time.
While the jet is active, 
we anticipate that the sum of the energy densities, 
$u_{\rm e} + u_{\rm m}$, scales as $\tau D^{-3}$,
since the total energy with which the jets supplied and the volume of the lobes 
are thought to be proportional to $\tau$ and $D^{3}$, respectively. 
The proportionality between $\tau$ and $D$ \citep{size2age} 
translate the relation into \bm{$u_{\rm e} + u_{\rm m} \propto D^{-2}$}.
The dispersion of jet power of FR II radio galaxies 
\citep[nearly an order of magnitude;][]{jet_power} may blur the relation. 
After the jets cease transporting sufficient energy to the lobes, 
a relation of $u_{\rm e} + u_{\rm m}\propto D^{-3} \exp(-t/T)$ is expected.
Here, $t$ is a time after the end of the jet activity, 
and $T \sim 0.2 $ Gyr $(1+z)^{-4} (1 + f)^{-1} (\gamma_{\rm e} / 10^{4})^{-1}$ 
is a cooling time scale of electrons due to synchrotron and IC radiation,
with $f$ being a ratio of \um~to the CMB energy density 
at the source rest frame, 
$u_{\rm CMB}(z) = 4.1 \times 10^{-13} (1+z)^{4}$ \edens.
For $t \ll T$, the radiation energy loss is neglected, 
and then, the lobe is thought to expand almost adiabatically  
to establish the pressure balance with its environment. 
In this second stage, the lobe is expected to follow 
a condition of \bm{$u_{\rm e} + u_{\rm m} \propto D^{-3}$}. 
On the other hand, in the final stage 
when the time-integrated energy loss become significant ($t \gg T$), 
we presume that \ue~and \um~decrease rapidly.
As a result, the lobe moves downward in the $D$--\ue~and $D$--\um~plots. 

The observed relations of 
$u_{\rm e} \propto D^{-2.2 \pm 0.4}$ and $u_{\rm m} \propto D^{-2.4 \pm 0.4}$ 
for the radio galaxies with $D < 1$ Mpc 
is close to that expected in the active phase of the jets
on the above argument.
The fact that \src~agrees with these correlations within a factor of $\sim 2$,
clearly shown in Figure \ref{fig:size2ue},
suggests that even the lobes of giant radio galaxies 
with a size of $D > 1$ Mpc are supplied with a sufficient energy
comparable to those into lobes with a moderate size ($D \sim 100$ kpc),
at least for the case of \src. 
However, this argument apparently conflicts 
with only a weak activity of the \src~nucleus,
which we suggested in this \suzaku~observation. 
In order to reconcile this inconsistency,
it is suggested that the \src~nucleus had been more active, 
and dimmed very recently.
As a result, it is implied 
that the rear end of the jet may have not yet arrived at the hot spot,
where the bulk jet energy is randomized and injected into the lobe. 
Adopting the typical speed of kpc-scale jets in FR II radio galaxies,
$0.5 c$ -- $0.7 c$ \citep{jet_speed},
it takes $\sim 5$ Myr for the jets to travel 
a distance of $\sim1$ Mpc from its nucleus. 
Therefore, we regard that it has been at most $5$ Myr
(neglecting the light travel time from the source to the Earth), 
since the nucleus of \src~declined its activity.
This means that the west lobe of \src~is about 
to make a transition into the second stage.

In Figure \ref{fig:size2ue},
we notice that one radio galaxy, Fornax A, 
deviates significantly below the trend of other radio galaxies
by more than an order of magnitude, especially in the \ue-$D$ plot. 
As mentioned in \S \ref{sec:nucleus}, 
\citet{ForA_nucleus} reported the dormancy of its nucleus,
based the \asca~observation.
They estimated that it has been $\sim 0.1$ Gyr,
since the end of its nuclear activity.
This is comparable to (or a few times) the electron life time in the lobes. 
All of these indicate that Fornax A has already entered the final stage.
   
We have just started to reveal the evolution of the energetics 
in lobes of radio galaxies, 
through IC X-ray observations on a large physical extent ($D>1$ Mpc).
However, our current knowledge is far from conclusive.
We regard it important to enlarge the X-ray sample 
of giant radio galaxies, and to fill the gap 
in the region of $D = 0.5$ -- $2$ Mpc in Figure \ref{fig:size2ue}.

\acknowledgments
We are grateful to all the members of the \suzaku~team,
for the successful operation and calibration.
We thank the anonymous referee for her/his great interest 
to the results and valuable comments to finalize the present paper.
Theoretical discussions with Dr. Kino improved the draft.
N. I. is supported by the Grant-in-Aid for the Global COE Program,
"The Next Generation of Physics, Spun from Universality and Emergence"
from the Ministry of Education, Culture, Sports,
Science and Technology (MEXT) of Japan.
P. G. acknowledges the RIKEN Foreign Postdoctoral Researcher Fellowship.
We have made extensive use of the NASA/IPAC Extra galactic Database
(NED; the Jet Propulsion Laboratory, California Institute
of Technology, the National Aeronautics and Space Administration).
The unpublished 1.4 GHz radio image of 3C 326 
was downloaded from ``An Atlas of DRAGNs'' 
\footnote{{\tt http://www.jb.man.ac.uk/atlas/}}, 
edited by Leahy, Bridle, \& Strom.      


\end{document}